# Strange Quark Star Model with Quadratic Equation of State

**Manuel Malaver**[*]

Universidad Marítima del Caribe, Departamento de Ciencias Básicas, Catia la Mar, Venezuela

*Corresponding author (Email:mmf.umc@gmail.com)

**Abstract -** In this paper, we studied the behaviour of compact relativistic objects with anisotropic matter distribution considering quadratic equation of state of Feroze and Siddiqui (2011). We specify the gravitational potential Z(x) in order to integrate the fields equations and there has been calculated the energy density, the radial pressure, the anisotropy and the mass function. The new solutions to the Einstein-Maxwell system of equations are found in term of elementary functions. For n=2, we have obtained the expressions for mass function, energy density, radius and metric functions of the model of Thirukkanesh and Ragel (2012) with polytropic equation of state.

## 1. Introduction

The study of the ultracompacts objects and the gravitational collapse is of fundamental importance in astrophysics and has attracted much since the formulation of the general theory of relativity. One of the fundamental problems in the general theory of relativity is finding exact solutions of the Einstein field equations [1,2]. Some solutions found fundamental applications ins astrophysics, cosmology and more recently in the developments inspired by string theory [2]. Different mathematical formulations that allow to solve Einstein´s field equations have been used to describe the behaviour of objects submitted to strong gravitational fields known as neutron stars, quasars and white dwarfs [3,4,5].

In the construction of the first theoretical models of relativistic stars are important the works of Schwarzschild [6], Tolman [7], Oppenheimer and Volkoff [8]. Schwarzschild [6] found analytical solutions that allowed describing a star with uniform density, Tolman [7] developed a method to find solutions of static spheres of fluid and Oppenheimer and Volkoff [8] used Tolman's solutions to study the gravitational balance of neutron stars. It is important to mention Chandrasekhar's contributions [9] in the model production of white dwarfs in presence of relativistic effects and the works of Baade and Zwicky [10] who propose the concept of neutron stars and identify a astronomic dense objects known as supernovas.

The description of the gravitational collapse and evolution of the compact objects has been a topic of great importance in general relativity. Many researchers have used a great variety of mathematical techniques to try to obtain exact solutions for relativistic stars, since it has been demonstrated by Komathiraj and Maharaj [4], Sharma et al. [5], Patel and Koppar [7], Tikekar and Singh [11], Feroze and Siddiqui [12], Malaver [13] and MafaTakisa and Maharaj [14]. These analyses indicate that the system of equations Einstein-Maxwell is important in the description ofultracompacts objects.

Komathiraj and Maharaj [15], Malaver [16], Bombaci [17], Thirukkanesh and Maharaj [18], Dey et al [19] and Usov [20] assume linear equation of state for quark stars. Feroze and Siddiqui [12] consider a quadratic equation of state for the matter distribution and specify particular forms for the gravitational potential and electric field intensity. MafaTakisa and Maharaj [14] obtained new exact solutions to the Einstein-Maxwell system of equations with a polytropic equation of state. Thirukkanesh and Ragel [21] have obtained particular models of anisotropic fluids with polytropic equation of state which are consistent with the reported experimental observations. More recently, Malaver [22] generated new exact solutions to the Einstein-Maxwell system considering Van der Waals modified equation of state with polytropic exponent. Mak and Harko [23] found a relativistic model of strange quark star with the suppositions of spherical symmetry and conformal Killing vector.

In this paper, we have studied of anisotropic fluids for uncharged matter with quadratic equation of state of Feroze and Siddiqui [12] in static spherically symmetric spacetime, using a similar procedure to the presented for Thirukkanesh and Ragel [21], where the gravitational potential depends on an adjustable parameter n. For a particular form of the potential, we have obtained a class of relativistic stars with the same values mass function, energy density, radius and metric functions, that those that it shows Thirukkanesh and Ragel [21] with polytropics indexes n=1 and n=2. This article is organized as follows, in Section 2, we present Einstein´s field equations. In Section 3, we make a particular choice of gravitational potential Z that allows solving the field equations. In Section 4, the physical features of the models are examined. Finally in Section 5, we conclude.

## 2. The field equations

Consider a spherically symmetric four dimensional spacetime whose line element is given in Schwarzschild coordinates by

$$ds^2 = -e^{2\nu(r)}dt^2 + e^{2\lambda(r)}dr^2 + r^2(d\theta^2 + \sin^2\theta d\varphi^2)  \tag{1}$$

$\nu(r)$ and $\lambda(r)$ are the two arbitrary functions. For uncharged perfect fluids the Einstein-Maxwell system of field equations are obtained as

$$\frac{1}{r^2}(1-e^{-2\lambda}) + \frac{2\lambda'}{r}e^{-2\lambda} = \rho \tag{2}$$

$$-\frac{1}{r^2}(1-e^{-2\lambda}) + \frac{2\nu'}{r}e^{-2\lambda} = p_r \tag{3}$$

$$e^{-2\lambda}\left(\nu'' + \nu'^2 + \frac{\nu'}{r} - \nu'\lambda' - \frac{\lambda'}{r}\right) = p_t \tag{4}$$

$\rho$ is the energy density, $p_r$ is the radial pressure and $p_t$ is the tangential pressure and primes denote differentiations with respect to r. Using the transformations suggested by Dugapal and Bannerji[24] $x = cr^2$, $Z(x) = e^{-2\lambda(r)}$ and $A^2 y^2(x) = e^{2\nu(r)}$ where A and c are arbitrary constants, the Einstein-Maxwell system has the equivalent form

$$\frac{1-Z}{x} - 2\dot{Z} = \frac{\rho}{c} \tag{5}$$

$$4Z\frac{\dot{y}}{y} - \frac{1-Z}{x} = \frac{p_r}{c} \tag{6}$$

$$4xZ\frac{\ddot{y}}{y} + (4Z + 2x\dot{Z})\frac{\dot{y}}{y} + \dot{Z} = \frac{p_t}{c} \tag{7}$$

where dots denote differentiation with respect to x.
With the transformations of[24], the mass within a radius r of the sphere take the form

$$M(x) = \frac{1}{4c^{3/2}}\int_0^x \sqrt{x}\rho(x)dx \tag{8}$$

In this paper, we assume the following quadratic equation of state

$$p_r = \alpha\rho^2 + \beta\rho - \gamma \tag{9}$$

proposed by Feroze and Siddiqui [12]. In eq. (9) $\alpha, \beta$ and $\gamma$ are arbitrary constants, $\rho$ is the energy density.

## 3. Classes of models

In this treatment, we have chosen the form of the gravitational potential $Z(x)$ as $Z(x) = (1-ax)^n$ where is a real constant and n is an adjustable parameter. This potential is well behaved and regular at the origin in the interior of the sphere. We have considered the particular cases for n=1,2, 3.

For case n=1, using $Z(x)$ in eq.(5), we obtain

$$\rho = 3ac \tag{10}$$

Substituting (10) in eq.(9), the radial pressure can be written in the form

$$p_r = 9a^2c^2\alpha + 3\beta ac - \gamma \tag{11}$$

Using (10) in (8), the expression of the mass function is

$$m(x) = \frac{a}{2\sqrt{c}} x^{3/2} \tag{12}$$

With (10) and (11), eq.(6) becomes

$$\frac{\dot{y}}{y} = \frac{9a^2c^2\alpha + 3\beta ac - \gamma + ac}{4c(1-ax)} \tag{13}$$

Integrating (13), we obtain

$$y(x) = c_1(-1+ax)^{\frac{(\gamma - a - 9\alpha c^2 a^2 - 3ca\beta)}{4a}} \tag{14}$$

$c_1$ is the constant of integration.

The anisotropy factor is defined as $\Delta = p_t - p_r$ and for n=1, $\Delta$ is given for

$$\Delta = \frac{xc(-\gamma + a + 9\alpha c^2 a^2 + 3ca\beta)(15c_1 + 2a)}{4(-1+ax)} \tag{15}$$

The metric functions $e^{2\lambda}$ and $e^{2\nu}$ can be written as

$$e^{2\lambda} = \frac{1}{1-ax} \tag{16}$$

$$e^{2\nu} = A^2 c_1^2 (-1+ax)^{\frac{(\gamma - a - 9\alpha c^2 a^2 - 3ac\beta)}{2a}} \tag{17}$$

With n=2, the expression for the energy density is

$$\rho = ac(6 - 5ax) \tag{18}$$

replacing (18) in (9), we have for the radial pressure

$$p_r = 25\alpha a^4 c^2 x^2 - (60\alpha a^3 c^2 + 5a^2\beta c)x + 36\alpha a^2 c^2 + 6\beta ac - \gamma \tag{19}$$

And the mass function is

$$m(x) = \frac{ax^{3/2}(2-ax)}{2\sqrt{c}} \tag{20}$$

The eq. (6) becomes

$$\frac{\dot{y}}{y} = \frac{4Ha^2 + (-25c^3a^3 + 50xc^2a^4)\alpha}{4a(-1+ax)} + \frac{(25xc^2a^3\alpha - 25x^2c^2a^4\alpha - F)}{4a(-1+ax)^2} \tag{21}$$

where for convenience we have let

$H = -\dfrac{5}{2a\alpha c^2} - \dfrac{5}{4\beta c} - \dfrac{1}{4}$ and $F = \gamma - a - c^2 a^2 \alpha - ca\beta$

Integrating eq. (21), we have

$$y(x) = c_2(-1+ax)^H e^{\frac{[-25xc^2a^3\alpha + 25x^2c^2a^4\alpha + F]}{4a(-1+ax)}} \tag{22}$$

$c_2$ is the constant of integration.

The metric functions $e^{2\lambda}$, $e^{2\nu}$ and the anisotropy factor $\Delta$ can be written as

$$e^{2\lambda} = \frac{1}{(1-ax)^2} \tag{23}$$

$$e^{2\nu} = A^2 c_2^2 (-1+ax)^{2H} e^{\frac{[-25xc^2a^3\alpha + 25x^2c^2a^4\alpha + F]}{2a(-1+ax)}} \tag{24}$$

$$\Delta = 4cx(1-ax)^2 \left\{ \begin{array}{l} \left[ \dfrac{Ha(H-1)}{(-1+ax)^2} + \left[ \dfrac{H}{4} - \dfrac{1}{2} \right] \left( \dfrac{-25c^2a^3\alpha + 50xc^2a^4\alpha}{(-1+ax)^2} \right) + \right. \\ \left( -\dfrac{Ha}{4} + \dfrac{1}{2} \right) \left( \dfrac{-25xc^2a^3\alpha + 25x^2c^2a^4\alpha + F}{(-1+ax)^3} \right) + \\ \left. \left[ \dfrac{1}{4} \left( \dfrac{-25c^2a^3\alpha + 50xc^2a^4\alpha}{a(-1+ax)} \right) - \dfrac{1}{4} \left( \dfrac{-25xc^2a^3\alpha + 25xc^2a^4\alpha + F}{(-1+ax)^2} \right) \right]^2 \end{array} \right\} - 2ac(1-ax) \left\{ 1 + 2x \left[ \dfrac{4Ha^2 + (-25c^3a^3 + 50xc^2a^4)\alpha}{4a(-1+ax)} - \left( \dfrac{-25xc^2a^3\alpha + 25x^2c^2a^4\alpha + F}{4(-1+ax)^2} \right) \right] \right\} + ac(2-ax) \quad (25)$$

With n=3, the expressions for $\rho$, $p_r$, $m(x)$, $y(x)$, $e^{2\lambda}$, $e^{2\nu}$ and $\Delta$ are given for

$$\rho = ac(9 - 15ax + 7a^2x^2) \tag{26}$$

$$p_r = 49\alpha c^2 a^6 x^4 - 210\alpha c^2 a^5 x^3 + (351\alpha c^2 a^4 + 7\beta ca^3)x^2 - (270\alpha a^3 c^2 + 15\beta a^2 c)x + 81a^2 c^2 \alpha + 9\beta ac - \gamma \tag{27}$$

$$m(x) = \dfrac{ax^{3/2}}{2\sqrt{c}} (3 - 3ax + a^2x^2) \tag{28}$$

$$e^{2\lambda} = \dfrac{1}{(1-ax)^3} \tag{29}$$

$$e^{2\nu} = A^2 c_3^2 (-1+ax)^{2E} e^{-\dfrac{\left[(49x^4 c^2 a^6 - 224x^3 c^2 a^5 + 301x^2 c^2 a^4)\alpha + Dx + G\right]}{4a(-1+ax)^2}} \tag{30}$$

where $D = 2a^2 + 2ca^2\beta - 122c^2a^3\alpha$, $G = \gamma - 3ca\beta - 5c^2a^2\alpha - 3a$ and

$$E = -\dfrac{1}{4} - \dfrac{7\beta c}{4} - \dfrac{15}{4}\alpha ac^2$$

$$\Delta = 4c(1-ax)^3 \left\{ \begin{array}{l} \dfrac{E^2 a^2 - Ea^2}{(-1+ax)^2} \\ + \dfrac{2Ea}{(-1+ax)} \left[ \begin{array}{l} -\dfrac{(196c^2a^6x^3\alpha - 672c^2a^5x^2\alpha + 602c^2a^4x\alpha + D)}{8a(-1+ax)^2} \\ + \dfrac{(49c^2a^6x^4 - 224c^2a^5x^3\alpha + 301c^2a^4x^2\alpha + Dx + G)}{4(-1+ax)^3} \end{array} \right] \\ - \dfrac{(588c^2a^6x^2\alpha - 1344c^2a^5x\alpha + 602c^2a^4\alpha)}{8a(-1+ax)^2} \\ - \dfrac{(196c^2a^6x^3\alpha - 672c^2a^5x^2\alpha + 602c^2a^4x\alpha + D)}{8(-1+ax)^3} \\ + \dfrac{3a}{32} \dfrac{(49c^2a^6x^4\alpha - 224c^2a^5x^3\alpha + 301c^2a^4x^2\alpha + Dx + G)}{(-1+ax)^4} \\ + \left( \begin{array}{l} -\dfrac{(196c^2a^6x^3\alpha - 672c^2a^5x^2\alpha + 602c^2a^4x\alpha + D)}{a(-1+ax)^2} \\ + \dfrac{(49c^2a^6x^4\alpha - 224c^2a^5x^3\alpha + 301c^2a^4x^2\alpha + Dx + G)}{4(1-ax)^3} \end{array} \right)^2 \end{array} \right\} - 3ac(1-ax)^2 \left\{ 1 + 2x \left[ \begin{array}{l} \dfrac{Ea}{(-1+ax)} - \dfrac{(196c^2a^6x^3\alpha - 672c^2a^5x^2\alpha + 602c^2a^4x\alpha + D)}{8a(-1+ax)^2} \\ + \dfrac{(49c^2a^6x^4\alpha - 224c^2a^5x^3\alpha + 301c^2a^4x^2\alpha + Dx + G)}{4(-1+ax)^3} \end{array} \right] \right\}$$

$$+ 3ac - 3a^2cx + a^3cx^2 \tag{31}$$

## 4. Physical features of the models

Any physically acceptable solutions must satisfy the following conditions [21,25]:
   (i) Regularity of the gravitational potentials in the origin.
   (ii) Radial pressure must be finite at the centre and it vanishes at the surface of the sphere.

   (iii) $p_r > 0$ and $\rho > 0$ in the origin.
   (iv) Decrease of the energy density and the radial pressure with the increase of the radius.

(v) In the surface of the sphere it should match with the Schwarzschild exterior solution, for which the metric is given by

$$ds^2 = -\left(1-\frac{2M}{r}\right)dt^2 + \left(1-\frac{2M}{r}\right)^{-1}dr^2 + r^2(d\theta^2 + \sin^2 d\phi^2) \quad (32)$$

With n=1, $e^{2\lambda(0)} = 1$, $e^{2\nu(0)} = A^2 c_1^2 (-1)^{\frac{(\gamma - a - 9\alpha c^2 a^2 - 3ac\beta)}{2a}}$ and

$\left(e^{2\lambda(r)}\right)'_{r=0} = \left(e^{2\nu(r)}\right)'_{r=0} = 0$. This demonstrates that the gravitational potential is regular in the origin.

In this model, the radial pressure is constant in the centre and the surface of the sphere, for what it is not a physically acceptable solution. In the origin and the boundary of the star.

$p_r(0) = p_r(r=R) = 9a^2c^2\alpha + 3\beta ac - \gamma$, $\rho(0) = \rho(r=R) = 3ac$, then $p_r$ and $\rho$ not satisfy (iv).

For the case n=2, $e^{2\lambda(0)} = 1$, $e^{2\nu(0)} = A^2 c_2^2 (-1)^{2H} e^{-\frac{F}{2a}}$ in the origin $r = 0$, $\left(e^{2\lambda(r)}\right)'_{r=0} = \left(e^{2\nu(r)}\right)'_{r=0} = 0$. This shows that the potential gravitational is regular in the origin.

In the centre $r = 0$, $\rho(0) = 6ac$ and $p_r(0) = 36\alpha a^2 c^2 + 6\beta ac - \gamma$, both are positive if a >0. In the surface of the star r=R, we have $p_r(r=R) = 0$ and $R = \sqrt{\frac{\beta + 12a\alpha c + \sqrt{\beta^2 + 4\alpha\gamma}}{10\alpha a^2 c^2}}$. If $\beta = \gamma = 0$, R take the form $R = \sqrt{\frac{6}{5ac}}$, expression deduced by Thirukkanesh and Ragel [21] with polytropic equation of state for polytropics indexes n=1 and n=2. Also it is calculated $\frac{d\rho}{dr} = -10a^2c^2 r < 0$ and $\frac{dp_r}{dr} = 100\alpha a^4 c^4 r^3 - 2cr(60\alpha a^3 c^2 + 5a^2 \beta c) < 0$ for all 0<r<R. The energy density and radial pressure decrease from the centre to the surface of the star.

From (20), we have

$$m(r) = \frac{1}{2}acr^3(2 - acr^3) \quad (33)$$

and the total mass of the star is

$$m(r=R) = \frac{ac}{2}\left[\frac{\beta + 12\alpha ac + \sqrt{\beta^2 + 4\alpha\gamma}}{10\alpha a^2 c^2}\right]^{3/2}\left[\frac{8\alpha a - \beta - \sqrt{\beta^2 + 4\alpha\gamma}}{10\alpha a}\right] \quad (34)$$

if $\beta = \gamma = 0$ eq. (34) take the form $m(r=R) = \frac{12\sqrt{6}}{25\sqrt{5ac}}$, also deduced by Thirukkanesh and Ragel [21].

Matching conditions for r=R can be written as

$$\left(1-\frac{2M}{R}\right) = A^2 y^2(cr^2) \quad (35)$$

$$\left(1-\frac{2M}{R}\right)^{-1} = \frac{1}{(1-acR^2)^2} \quad (36)$$

To maintain of causality, the radial sound speed defined as $v^2_{sr} = \frac{dp_r}{d\rho}$ should be within the limit $0 \leq v^2_{sr} \leq 1$ in the interior of the star. In this model, we have

$v^2_{sr} = \frac{dp_r}{d\rho} = 2\alpha ac(6 - 5acx) + \beta$, and so we must impose the condition

$0 \leq 12\alpha ac - 10\alpha a^2 c^3 r^2 + \beta \leq 1$

With n = 3, $e^{2\lambda(0)} = 1$, $e^{2\nu(0)} = A^2 c_3^2 (-1)^{2E} e^{-\frac{G}{2a}}$ in the origin and

$\left(e^{2\lambda(r)}\right)'_{r=0} = \left(e^{2\nu(r)}\right)'_{r=0} = 0$. Again the gravitational potential is regular in $r = 0$.

In the centre $\rho(0) = 9ac$ and $p_r(0) = 81\alpha a^2 c^2 + 9\beta ac - \gamma$, both are positive if a >0. In the boundary of the star r=R, we have $p_r(r=R) = 0$ and $R = \sqrt{\frac{30a\alpha c + 2\sqrt{-27\alpha^2 c^2 a^2 - 14ca\beta + 14\alpha ca\sqrt{\beta^2 + 4\alpha\gamma}}}{28\alpha a^2 c^2}}$ If $\Delta = -27\alpha^2 c^2 a^2 - 14ca\beta + 14\alpha ca\sqrt{\beta^2 + 14\alpha\gamma} = 0$, we obtain $R = \sqrt{\dfrac{15}{14ac}}$. This is a new value found for the radius of the star.

Then, $\dfrac{d\rho}{dr} = 2a^2 c^2(-15r + 14acr^3) < 0$ and

$\dfrac{dp_r}{dr} = 392\alpha c^6 a^6 r^7 - 1260\alpha c^5 a^5 r^5 + (1404\alpha c^4 a^4 + 28\beta c^3 a^3)r^3 - (540\alpha a^3 c^3 + 30\beta c^2 a^2) < 0$

for all 0<r<R. The energy density and radial pressure decrease from the centre to the surface of the star.

From (28), we have

$$m(r) = \frac{1}{2} acr^3 (3 - 3acr^2 + a^2 c^2 r^4) \tag{37}$$

and the total mass of the star is

$$m(r = R) = \frac{ac}{2} \left[ \frac{30\alpha ac + 2\sqrt{-27\alpha^2 c^2 a^2 - 14\alpha ca\beta + 14\alpha ca\sqrt{\beta^2 + 4\alpha\gamma}}}{28\alpha a^2 c^2} \right]^{3/2}$$

$$\left[ 3 - 3ac \left( \frac{30\alpha ca + 2\sqrt{-27\alpha^2 c^2 a^2 - 14\alpha ca\beta + 14\alpha ca\sqrt{\beta^2 + 4\alpha\gamma}}}{28\alpha c^2 a^2} \right) \right.$$

$$\left. + a^2 c^2 \left( \frac{30\alpha ca + 2\sqrt{-27\alpha^2 c^2 a^2 - 14\alpha ca\beta + 14\alpha ca\sqrt{\beta^2 + 4\alpha\gamma}}}{28\alpha c^2 a^2} \right)^2 \right] \tag{38}$$

if $\Delta = 0$ eq. (38) take the form $m(r = R) = \dfrac{2745\sqrt{15}}{5488\sqrt{14ac}}$. Matching conditions for r=R can be written as

$\left(1 - \dfrac{2M}{R}\right) = A^2 y^2(cr^2)$ and $\left(1 - \dfrac{2M}{R}\right)^{-1} = \dfrac{1}{(1 - acR^2)^3}$

For this case, the condition $0 \leq v^2_{sr} \leq 1$, implies that $0 \leq 18\alpha ac - 30\alpha a^2 c^2 r^2 + 14\alpha a^3 c^3 r^4 + \beta \leq 1$

## 5. Conclusion

In this paper, we have found a class of models with quadratic equation of state for the radial pressure that correspond to anisotropic compact sphere, where the gravitational potential Z depends on an adjustable parameter n. All the generated models for n>1, satisfy the physical characteristics of a realistic star. For the case n=2, we obtain the same expressions for mass function, energy density, radius of the star and metric function $e^{2\lambda}$ of the model of Thirukkanesh and Ragel [21] with polytropics indexes n=1 and n=2.

For n=3, we have obtained a new model of star where the radius R and the total mass of the star $(M)$ is given for $R = \sqrt{\dfrac{15}{14ac}}$ and $M = \dfrac{2745\sqrt{15}}{5488\sqrt{14ac}}$. This solution can be used to describe compact objects. Following Thirukkanesh and Ragel [17], if the parameter a=0.024 km$^{-2}$ and c=1, we obtain for R=6.7 km and a=0.0071 km$^{-2}$, c=1 we obtain R=12.3 km. Experimental observations show the existence of compact objects with radius of 7km and 13km, respectively [26,27]. Therefore, with the models generated in this article we can obtain stars of mass and densities comparable to the experimental results.